\documentclass[%
 reprint,
 amsmath,amssymb,
 aps,
]{revtex4-2}
\usepackage{upgreek}
\usepackage{float}
\usepackage{color}
\usepackage{braket}
\usepackage{hyperref}
\usepackage{url}
\usepackage{graphicx}% Include figure files
\usepackage{epstopdf}
\usepackage{dcolumn}% Align table columns on decimal point
\usepackage{bm}% bold math
\hypersetup{colorlinks=true, urlcolor=blue, citecolor=blue}
\begin{document}

\preprint{APS/123-QED}

\title{Pressure-induced Superconductivity in Tellurium Single Crystal}% Force line breaks with \\

\author{Lingxiao Zhao$^1$}
\author{Cuiying Pei$^1$}
\author{Juefei Wu$^1$}
\author{Yi Zhao$^1$}
\author{Qi Wang$^{1,2}$}
\author{Bangshuai Zhu$^1$}
\author{Changhua Li$^1$}
\author{Weizheng Cao$^1$}
\author{Yanpeng Qi$^{1,2,3}$}
    \altaffiliation[Correspondence ]{should be addressed to Yanpeng Qi (qiyp@shanghaitech.edu.cn)}
    
\affiliation{$^1$School of Physical Science and Technology, ShanghaiTech University, Shanghai, China\\
$^2$ShanghaiTech Laboratory for Topological Physics, ShanghaiTech University, Shanghai, China\\
$^3$Shanghai Key Laboratory of High-resolution Electron Microscopy, ShanghaiTech University, Shanghai, China}%Lines break automatically or can be forced with \\

\date{\today}% It is always \today, today,

\begin{abstract}
	Tellurium (Te) is one of the $p$-orbital chalcogens, which shows attractive physical properties at ambient pressure. Here, we systematically investigate both structural and electronic evolution of Te single crystal under high pressure up to 40 GPa. The pressure dependence of the experimental Raman spectrum reveals the occurrence of multiple phase transitions, which is consistent with high-pressure synchrotron X-ray diffraction (XRD) measurements. The appearance of superconductivity in high-pressure phase of Te is accompanied by structural phase transitions. The high-pressure phases of Te reveal a nonmonotonic evolution of superconducting temperature $T_c$ with notably different upper critical fields. The theoretical calculations demonstrate that the pressure dependence of the density of states(DOS) agrees well with the variation of $T_c$. Our results provide a systematic phase diagram for the pressure-induced superconductivity of Te.
	
\end{abstract}

\maketitle

%\tableofcontents

\section{Introduction}

Group VI tellurium (Te) belongs to the chalcogen element family and is a very interesting element in terms of electronic properties. At ambient pressure, Te has a trigonal crystal structure (Te-I structure) with the $P3_121$ or $P3_221$ space group, depending on the sense of rotation of the helical chains (right- or left-handed screw)\cite{hirayama2015weyl}. Each Te atom is covalently bonded with its two nearest neighbors on the same chain, and the interchain interactions are van der Waals-like bonds that are weaker than the covalent ones. The lone-pair and anti-bonding orbitals give rise to a slightly indirect bandgap about 0.35 eV, which has the conduction band minimum (CBM) located at the H point of the Brillouin zone, and the valence band maximum (VBM) that is slightly shifted from the H point along the chain direction\cite{deng2006unusual,hirayama2015weyl}. Although its rarity in the earth's crust is comparable to that of platinum, Te possesses multifunctional properties, e.g. semiconductivity, piezoelectricity, photoconductivity, thermoelectricity and topological insulator, for applications in sensors, energy devices, optoelectronics, and electronics\cite{RN40,xian2017square,liu2010mesostructured,RN42,RN36,RN38,RN35,RN33,RN44,RN34}.\\
\indent The inherent anisotropy of one dimensional(1D) chiral chains in Te makes it very sensitive to external pressures. At room temperature, Te shows multiple structural phase transitions by varying the applied pressure: ambient phase Te-I (trigonal) to Te-II (triclinic) at around 4 GPa, Te-II to Te-III (modulated body centered monoclinic) at around 8.5 GPa and Te-III to Te-V (body centered cubic) at around 30 GPa\cite{aoki1980crystal,RN7,RN5,RN21,RN4,RN6}. Since Te-III and Te-II usually coexist together in the pressure range from 4.5 to 8.5 GPa, the boundary between the two phases is not specific\cite{RN21}. In addition, Te-IV (rhombohedral) phase only exists at high pressure (about 27 GPa) and high temperature (over 300 K), which is not relevant to low-temperature properties (e.g. superconductivity).\cite{RN21}\\
\indent Application of pressure dramatically alters not only the crystal structure but also the electronic properties in Te. The pressure-induced topological phase transition from a semiconductor to a Weyl semimetal was reported in Te-I at a quite low-pressure region\cite{agapito2013novel,hirayama2015weyl,deng2006unusual}. More interestingly, all other phases except for the Te-I and Te-IV phases mentioned above were reported to show superconductivity in low temperature\cite{mauri1996phonon,akahama1992pressure,RN10}. In addition, Te is also a crucial component of many functional materials, e.g., tellurides, some of them show superconductivity at ambient pressure\cite{ishihara1983electrical,sales2009bulk,hanaguri2010unconventional,liu2016nature,correa2023superconductivity,hein1970superconductivity,qi2016superconductivity}. Although these pieces of knowledge have already been accumulated since 1960s\cite{RN10,akahama1992pressure}, the detailed pressure-dependent evolution of superconductivity and magnetic properties of Te single crystals remains little known in the literature. Thus, it is necessary to systematically investigate the basic superconducting properties of Te under high pressure.\\
\indent In this paper, we report the pressure-dependent Raman spectroscopy and transport properties of Te single crystal under pressures up to 40 GPa. A nonmonotonic evolution of $T_c$ is observed, accompanied by multiple structural phase transitions. The superconducting phase diagram of Te under high pressures is also obtained. Our results reveal the interesting and rich physics in Te single crystals under high pressure. In addition, pressure-induced superconductivity was reported in some typical tellurides recently\cite{zhang2011pressure,MEinaga_2010,qi2016pressure,liu2017superconductivity,qi2017topological,zhao2022pressure,pei2022pressure,PhysRevMaterials.6.L051801,PhysRevB.103.174515,Chen_2023,Zhu2023}. It should be mentioned that cautions must be taken when claiming superconductivity of tellurides under pressure since Te impurity might exist in the sample. We hope that our study will also serve as a useful reference for this purpose.

\begin{table}[H]%The best place to locate the table environment is directly after its first reference in text
    
    \caption{\label{ST}High-pressure phases of Te. Bcm and bcc stand for body centered monoclinic and body centered cubic, respectively. IM means the incommensurate modulated nature of Te-III.%
    }
    \begin{ruledtabular}
    \begin{tabular}{cccc}
    \textrm{Pressure range}&
    \textrm{Phase}&
    \textrm{Bravais Lattice}&
    \textrm{Spacegroup}
    \\
    
    \colrule
    0 - 4 GPa & Te-I & trigonal & $P$3$_1$21 \\
    4 - 8.5 GPa & Te-II & triclinic & $P$-1 \\
    8.5 - 30 GPa & Te-III & bcm (IM) & $C$2/m \\
    above 30 GPa & Te-V & bcc & $I$m-3m \\
    \end{tabular}
    \end{ruledtabular}
\end{table}

\section{Methods}
Single crystals are picked from the commercial product of Te powder by Alfa Aesar (99.999\% purity), with typical dimensions of 0.3 mm × 0.05 mm × 0.01 mm. The single crystals have glittering surfaces. The sample quality is confirmed by single crystal XRD and the details are shown in Table. S1.\\
\indent High-pressure $in$-$situ$ Raman spectroscopy is conducted on a Raman spectrometer (Renishaw in Via, U.K.). The pressure is implemented by a symmetric Diamond Anvil Cells (DAC) with 300$\upmu$m-culet diamonds. Daphne oil 7373 is utilized as the pressure-transmitting medium. \\
\indent The in-field resistivity experiments are conducted with a Physical Property Measuring System. The resistivity measurements without a magnetic field are carried out with a cryogenic measuring system. The resistivity is measured by van der Pauw four-probe method. To collect a detailed phase diagram of resistivity under pressure, diamonds with 300/400 $\upmu$m culets are used, to implement pressures up to 40 GPa. Cubic Boron Nitride (cBN) is utilized as the insulating layer.\cite{10.1093/nsr/nwad034,10.1063/5.0088235} Thin Platinum plates are cut into needle-like electrodes to connect the sample in the pressure chamber with the Copper lines outside the chamber.\\
\indent The DC magnetic susceptibility measurements at high pressures are accomplished by using the DAC with 500$\upmu$m-culet diamonds. The DAC without sample is firstly measured for the background signals. In order to obtain high-quality signals, we put as many samples as possible into the sample chamber without pressure transmitting medium. Magnetization measurements are carried out in zero-field-cooling (ZFC) and field-cooling (FC) modes under 20 Oe in the low temperature region. In all the high-pressure experiments in this research, the fluorescence of ruby is used for calibrating the scale of pressure.\cite{Mao}\\

\indent  $Ab$-$initio$ calculations are conducted with the Vienna $ab$-$initio$ simulation package (VASP) within the framework of density functional theory (DFT)\cite{PhysRevB.47.558,PhysRevB.54.11169,GKresse_1994}. The Perdew-Burke-Erzernhof (PBE) functional based on the generalized gradient approximation (GGA) is chosen to describe the exchange-correlation interaction, and the projector augmented wave (PAW) method is adopted with the energy cutoff of plane-wave basis set at 400 eV. The convergence criteria for geometry optimization and atomic relaxation are set 0.003 eV/$\rm \AA$ and 10$^{-6}$ eV per atom for force and energy, respectively. A Monkhorst-Pack $k$-point grids with a reciprocal spacing $2\pi \times 0.03 {\rm \AA}^{-1}$ in the Brillouin zone is selected. 

\begin{figure*}[t]
    \centering
    \includegraphics{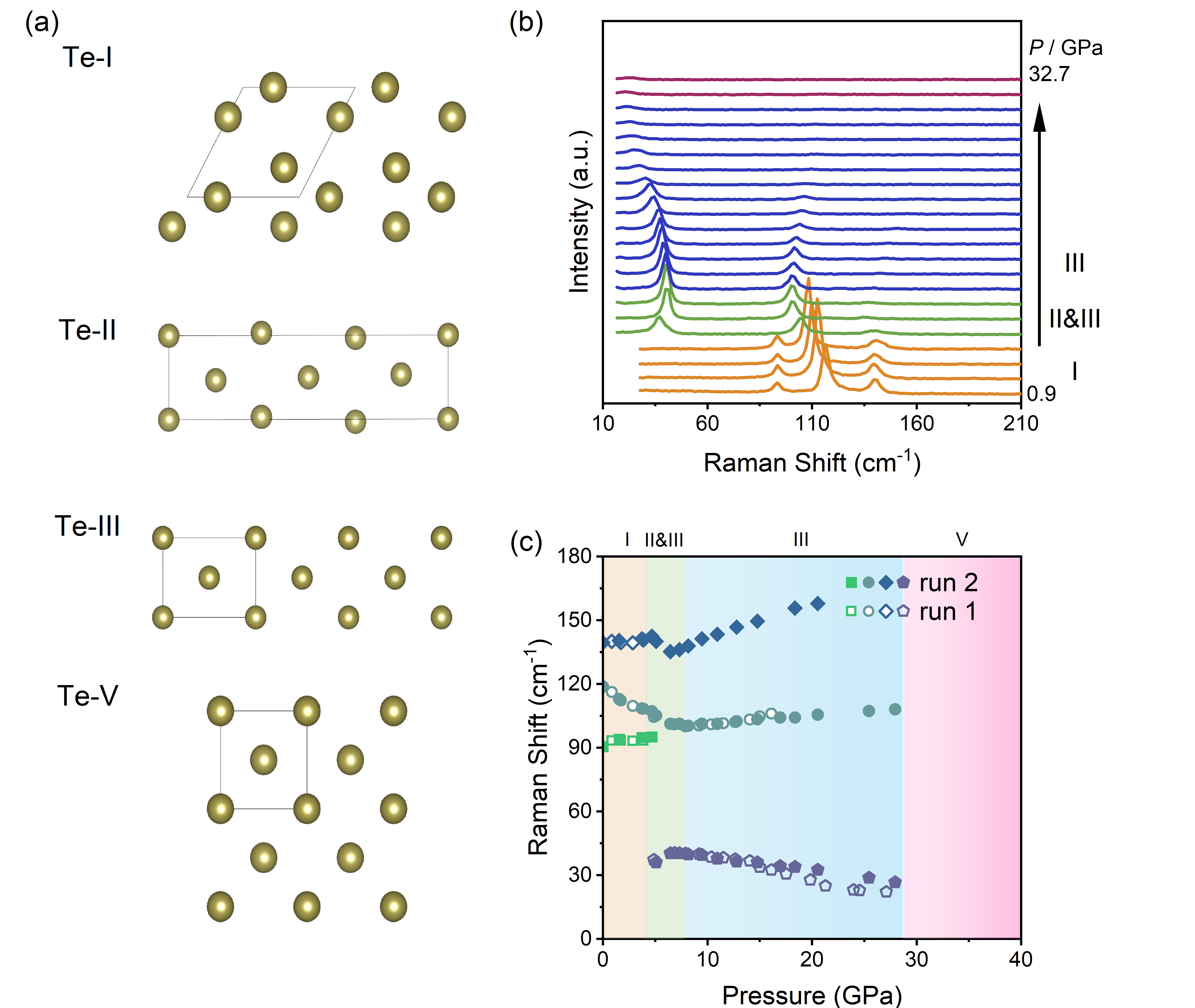}
    %\captionsetup{justification=raggedright}   
    \caption{The crystal structure and Raman spectroscopy of Te under high pressures. (a) The crystal structures of Te-I, Te-II, Te-III, Te-V with $c$ axis. (b) Pressure dependence of Raman spectra under various pressures for Te at room temperature. (c) Raman shifts of Te in compression. The colored backgrounds show different phases of Te, as summarized in Table. I.}
    \label{raman}
\end{figure*}
\begin{figure*}[t]
    \centering
    \includegraphics{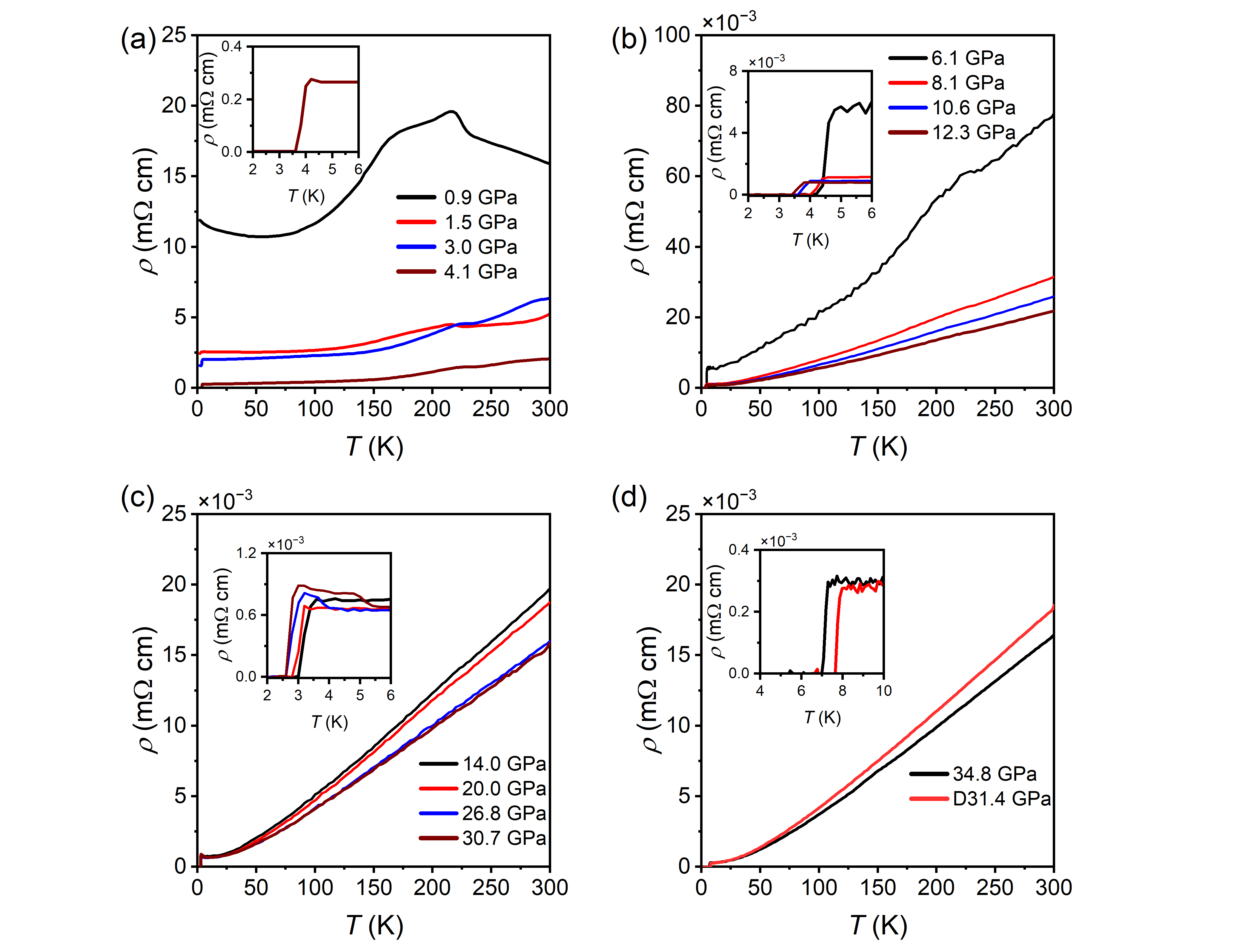}
    %\captionsetup{justification=raggedright}   
    \caption{Electrical resistivity of Te as a function of temperature for various pressures. The insets of each figures show temperature-dependent resistivity of Te in the vicinity of the superconducting transition. The red line in (d) shows the resistivity data under decompression to 31.4 GPa.  }
    \label{ttRT}
\end{figure*}
\begin{figure*}[t]
    \centering
    \includegraphics{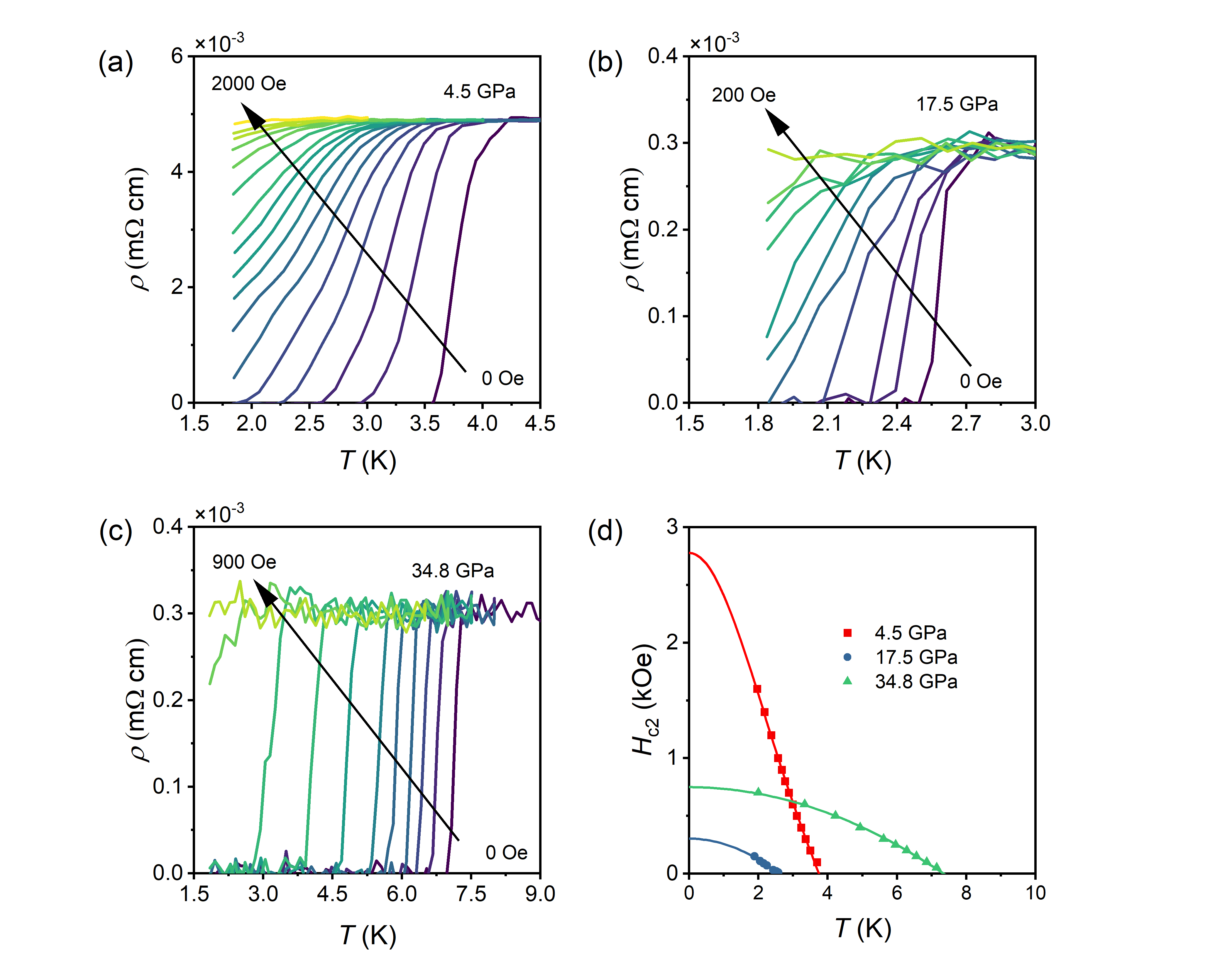}
    %\captionsetup{justification=raggedright}   
    \caption{Temperature dependence of resistivity under different magnetic fields for Te at 4.5 GPa (a), 17.5 GPa (b) and 34.8 GPa (c), respectively. (d) Estimated upper critical field for Te. Here, $T_c$ is determined as a 90\% drop in the normal-state resistivity. The solid lines represent fits based on the formula (\ref{GGL}).}
    \label{ifRT}
\end{figure*}
\begin{figure*}[t]
    \centering
    \includegraphics{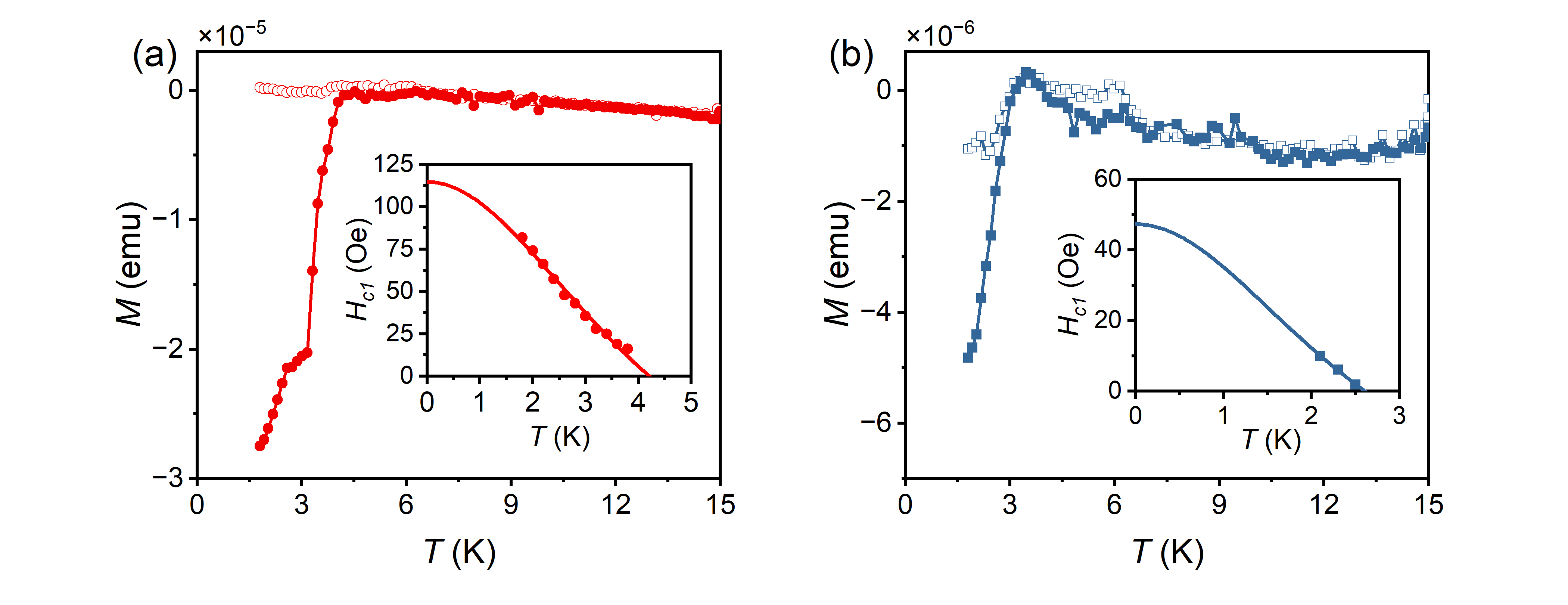} 
    %\captionsetup{justification=raggedright} 
    \caption{$M-T$ curves at 6.0 GPa (a) and at 8.9 GPa (b). The hollow symbols show the FC curves and the solid symbols show the ZFC curves. The insets show the fitted results of $H_{c1}$ from $M$-$H$ measurements. }
    \label{MPMS}
\end{figure*}
\begin{figure*}[t]
    \centering
    \includegraphics{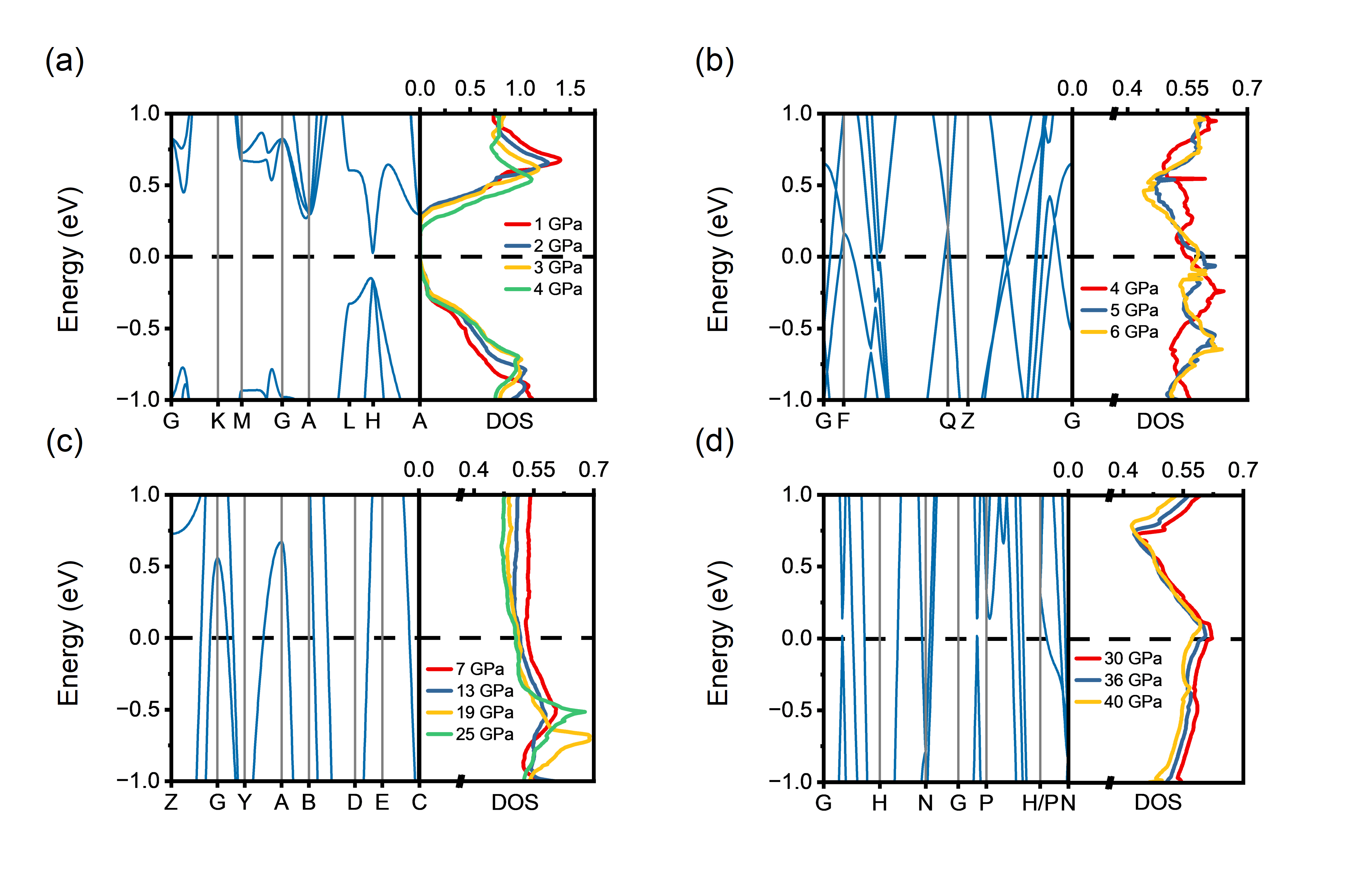} 
    %\captionsetup{justification=raggedright}  
    \caption{Electronic band structure and density of states (DOS) of Te at various pressures. The electronic structures of the four phases from DFT results are shown. In each figure, the left graphs show the band structure and the right graphs show the DOS, with the vertical axis representing energy in $E-E_F$ and the horizontal axis representing DOS per atom in a unit cell.}
    \label{DFT}
\end{figure*}

\section{Results and Discussions}
\indent It has been previously reported that Te crystal at room temperature sequentially undergoes multiple structural phases with applying pressure. As shown in Fig. \ref{raman}(a), Te-I has a trigonal structure, while Te-II, Te-III and Te-V show body-centered structures. Though the symmetries of the latter three phases are notably different, the occupied wyckoff positions of Te atoms are quite similar\cite{RN7,RN4,RN6}. Raman spectroscopy is a powerful tool to probe changes in the crystal lattice, and thus, our pressure-dependent Raman spectra of Te are accompanied by electronic transport measurements at various pressures shown below. Fig. 1(b) shows the Raman spectra of Te single crystal at various pressures. At 0.9 GPa, three peaks are assigned as follows: E(1) = 90 cm$^{-1}$, A1 = 120 cm$^{-1}$ and E(2) = 140 cm$^{-1}$. As the pressure increases, we can distinguish three phase transitions. In the pressure range of 0 - 4 GPa, the profile of the spectra remains similar to that at ambient pressure, and the observed E vibrational modes exhibit blue-shift tendencies, which is the typical behavior under high pressure, while the A1 mode displays the opposite trend and shows red-shift behavior.\\
\indent At the narrow pressure range (around 4 - 5 GPa), a new peak Ag(1) = 37 cm$^{-1}$ appears, indicating that Te-I transforms to Te-II.  It should be noted that Ag(1) peak shown here is consistent with $ab$-$initio$ theoretical prediction\cite{RN5,C7CP08002F}. The other peaks are the Ag(2) mode at 105 cm$^{-1}$ and Ag(3) mode at 140 cm$^{-1}$. A pressure-induced phonon softening is observed in Ag(3) mode in this pressure range, due to the instability of Te-II\cite{RN5}.\\
\indent High-pressure $in$-$situ$ XRD demonstrated that Te-II is only stable in a narrow pressure range and coexists with Te-III upon compression\cite{RN7,RN21}. Above 8.5 GPa, only pure Te-III exists. The reported pristine Te-III structure has a $C$2/m space group, with Te atoms occupying two wyckoff points, (0, 0, 0) and (0.5, 0.5, 0)\cite{RN7}. According to group theory analysis, however, all the phonon modes at $\Gamma$ are Raman inactive ones. Recent XRD measurements have provided evidence for the appearance of an incommensurately modulated (IM) lattice arrangement\cite{RN7}. The IM structure in Te-III is caused by charge density waves (CDW), with a modulation wave vector $q$ = (0, 0.288, 0)\cite{RN7}. Thus, the observed Raman peaks after 8.5 GPa originate from the incommensurate modulation (IM) of the Te-III structure. Noted that the observed strong peak of Ag(1) as well as phonon softening behavior present further evidence of the structural modulations.\\
\indent Upon further compression exceeding a pressure of 30 GPa, all the peaks disappear, indicating a structural phase transition from Te-III to Te-V. In addition, a reversible phase transition associated with a compressed lattice is verified by the Raman spectrum of the sample after recovery to 1 atm, as shown in Fig. S1\cite{SI}. In summary, our Raman study is consistent with previous synchrotron XRD measurements and provides further evidence for pressure-induced multiple structural phase transitions of Te single crystals. The details of phase transitions under high pressure are summarized in Table I.\\   
\indent Temperature dependence of resistivity $\rho(T)$ of Te under pressure up to 34.8 GPa is shown in Fig. \ref{ttRT}. At ambient pressure, Te is a $p$-type semiconductor with a narrow bandgap, while the resistivity alters dramatically with increasing applied pressure. At 0.9 GPa, the $\rho(T)$ first increases with decreasing temperature and reaches a maximum value at around 200 K. Then the $\rho(T)$ gradually decreases showing metallic behavior with a positive d$\rho(T)$/dT slope. Similar anomaly was also observed in previous research, which is caused by the impurity energy levels\cite{RN14}.  Upon further increasing the pressure, the resistivity begins to drop rapidly and the semiconducting-like behavior is suppressed. At 1.5 GPa, the $\rho(T)$ shows typical metallic behavior in the whole temperature region. When the pressure is increased to 3.0 GPa, a tiny drop of resistivity is observed at the lowest temperature. Further increasing pressure to 4.1 GPa, superconductivity occurs with critical temperature $T_c$ of 4.0 K.  According to the structural phase diagram of Te under high pressures\cite{aoki1980crystal,RN7}, the Te-I phase transforms to Te-II at about 4 GPa, therefore we can probably regard Te-II as the superconducting phase with $T_c$ = 4.0 K, so the superconductivity induced by pressure is closely associated with structural phase transitions. At the narrow pressure range of Te-II, the superconductivity is robust and changes slowly, while $T_c$ starts to decrease above 8.1 GPa in Te-III and is suppressed to a minimum of 2.8 K at 30.7 GPa. When external pressure increases to 34.8 GPa, a superconducting phase with higher $T_c$ of 7.2 K appears, where Te-III transforms to Te-V. $T_c$ starts to decrease monotonically with further increasing pressure in Te-V. Note that $T_c$ further increases under decompression and reaches a maximum value of 8 K at 31.4 GPa.\\

\begin{table}[H]%The best place to locate the table environment is directly after its first reference in text
    \caption{\label{Hc2}The superconducting properties of Te at various pressures.   %
    }
    \begin{ruledtabular}
    \begin{tabular}{ccccccc}
    
    \textrm{$P$ (GPa)}&
    \textrm{Phase}&
    \textrm{$T_c$ (K)}&
    \textrm{$H_{c2}$ (Oe)}&
    $\alpha$ &
    $\beta$ &
    $\xi(0)$ (nm)\\
    \colrule
    4.5 & Te-II & 4.0 &2778 & 1.00 & 1.00 & 34.4 \\
    17.5 & Te-III & 2.8 & 303 & 1.00 & 0.00 & 104.3\\
    34.8 & Te-V & 7.2 & 748 & 1.00 & 0.00 & 66.4\\
    \end{tabular}
    \end{ruledtabular}
\end{table}

\indent To gain insights into the superconducting transition, we measure the resistivity at different magnetic fields to obtain the upper critical fields $H_{c2} $(0) of each phase. Fig. \ref{ifRT} shows $\rho(T)$ under different magnetic fields at 4.5 GPa, 17.5 GPa and 34.8 GPa, respectively. When increasing the magnetic field, the resistivity drop is continuously shifted to a lower temperature. These results indicate that the sharp drop in resistivity is a superconducting transition. The Te-II phase has a relatively higher $H_{c2} $(0). A field of 2000 Oe can almost suppress the superconductivity completely above 1.8 K. In sharp contrast, the Te-V phase has a much lower $H_{c2}$(0), although the $T_c$ of Te-V is almost two times higher than that of Te-II. We extract the field dependence of $T_c$ for Te at different pressures and plot $H(T)$ in Fig. \ref{ifRT}(d). The slopes of $dH_{c2}/dT$ are notably different under various pressures, indicating discriminate superconducting mechanisms of each phase. \\

\indent We fit $H(T)$ curves using the simple formular for each phase as follows\cite{RN31,RN12}. \\
\begin{equation}
    H_{c2}(T)=H_{c2}(0) \times \frac{[1-(T/T_c)^2]^\alpha}{[1+(T/T_c)^2]^\beta}
    \label{GGL}
\end{equation}
\indent In equation (1), the both temperature-dependent terms $ (1-(T/T_c)^2)^\alpha $ and $ (1+(T/T_c)^2)^{-\beta} $ are taken into consideration. The fitting curves are indicated by the colored solid lines in Fig. 3(d). As we can see, Te-II has an upper critical field of 2778 Oe. This upper critical field is higher than some other superconducting pure elements with similar $T_c$, like tin (3.72 K, 308 Oe), indium (3.40 K, 286 Oe), and tantalum (4.48 K, 830 Oe). Both Te-III and Te-V have a very low upper critical field of 303 Oe and 748 Oe, respectively. For type-II superconductors in the dirty limit, a simple estimate using the conventional one-band Werthamer-Helfand-Hohenberg (WHH) approximation $ H_{c2}(0) = 0.691 \times (dH_{c2}/dT) \times T_c $, yields a value of 2101 Oe, 291 Oe, 861 Oe for Te-II, Te-III and Te-V, respectively. These upper critical fields are much lower than the Pauli limiting fields, $H_{P}(0) = 1.84T_c$, respectively, indicating that Pauli pair breaking is not relevant. According to the relationship between $H_{c2}$(0) and the coherence length $\xi$(0), namely, $ H_{c2}(0) = \Phi_0/(2\pi \xi(0) ^2) $, where $ \Phi_0 = 2.07 \times 10^{-15} $ Wb is the flux quantum, the derived $\xi$(0) could be obtained. The corresponding data are summarized in Table. \ref{Hc2}. The up
er critical field is a fundamental measure of the strength of superconductivity in a material. Further research is needed to clarify why the different phases of Te have such different upper critical fields.\\

\indent In order to further confirm the superconductivity of Te under high pressure, a DC susceptibility measurement is also performed. As shown in Fig. 4, a sharp transition with a large diamagnetic signal can be clearly seen at 4.0 K under 6.0 GPa, which corresponds to the superconducting transition of phase Te-II. Meanwhile, a transition with a tiny diamagnetic signal is also visible at about 2.8 K, which may reflect the superconducting transition corresponding to Te-III. When we increase the pressure to 8.9 GPa, only the diamagnetic signal of Te-III is observed. Then, at 8.9 GPa, Te-III becomes the dominant superconducting phase. The magnetization curves in the superconducting states show the typical behavior of type-II superconductors.\\
\indent The magnetization versus external field over a range of temperatures below $T_c$ is presented in Fig. S2\cite{SI}. The field deviating from a linear curve of full Meissner effect is deemed as the lower critical field $H_{c1}$ at each temperature and is summarized in the insets of Fig. 4. The $H_{c1}(0)$ data points can be well fitted with a simple formula.\\
\begin{equation}
    H_{c1}(T)=H_{c1}(0) \times \frac{1-(T/T_c)^2}{1+(T/T_c)^2}
    \label{HC1}
\end{equation}
\indent The obtained $H_{c1}(0)$ are 115 Oe for Te-II at 6.0 GPa and 47 Oe for Te-III at 8.9 GPa, respectively.

\indent To theoretically understand the evolution of physical properties under pressure, we have performed density functional theory (DFT) calculations on different phases of Te at 0 GPa, 4 GPa, 13 GPa and 30 GPa, respectively. The lattice parameters and atom positions are fully relaxed under various pressures, and the optimized cell volumes and shapes at different pressures and of different phases agree well with the experimental refinement results as shown in Fig. S3\cite{SI}. The band structures and DOS of Te with different phases are shown in Fig. \ref{DFT}. Te-I is a normal semiconductor with a narrow band gap (E$_g$ = 0.16 eV) at ambient pressure. With increasing pressure, the band gap undergoes a nonmonotonic process, which is closely related to a pressure-induced topological transition of Te-I at around 2-3 GPa\cite{agapito2013novel,hirayama2015weyl,deng2006unusual}. With the further increase of pressure, three structural phase transitions are observed for Te. Different from the ambient phases, all high-pressure phases are metallic and exhibit a finite DOS at $E_F$, which is consistent with our resistivity data. The complicated modulated structure of Te-III makes phonon dispersions and electron-phonon interaction calculations very time-consuming, herein, the electronic structure of unmodulated bcm structure is calculated. As shown in Fig. \ref{DFT}(b), the band structure of Te-III still shows a metallic behavior. Since multiple bands cross the Fermi level, there exists a Fermi surface nesting in Te-III, which is probably the origin of the CDW phase\cite{PhysRevLett.102.035501}.\\

\begin{figure}[H]
    \centering
    \includegraphics{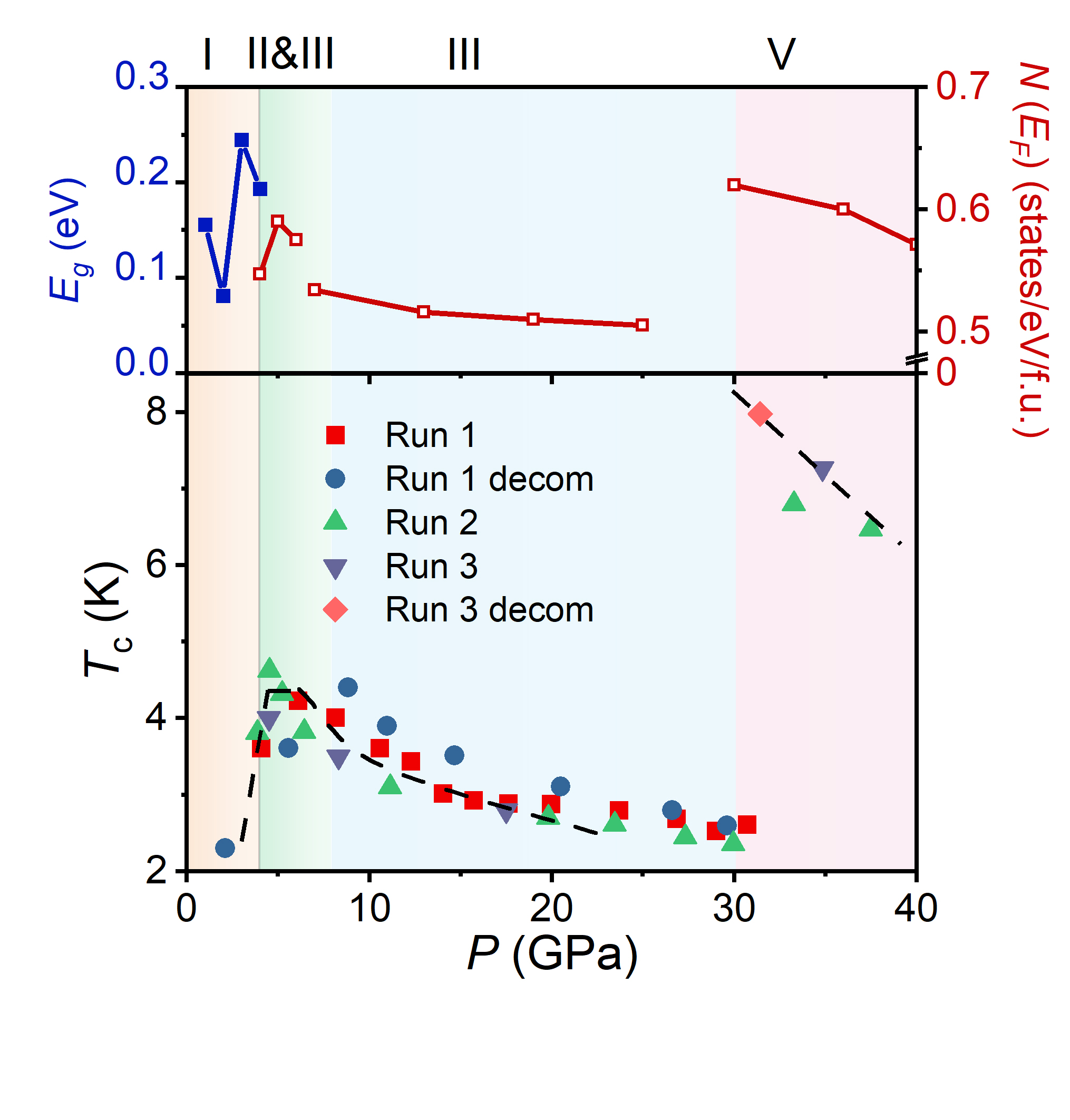} 
    %\captionsetup{justification=raggedright} 
    \caption{Electronic phase diagram for Te under pressures. The colored areas represent different phases. The upper panel shows the pressure dependence of the calculated band gaps of Te-I, and the density of states (DOS) at the Fermi level for Te-II, Te-III and Te-V. The lower panel shows the superconducting $T_c$ as a function of pressure. The solid symbols represent the $T_c$ extracted from different runs of resistivity measurements.}
    \label{pd}
\end{figure}

\indent Several independent high-pressure transport measurements on Te single crystals provide consistent and reproducible results, as shown in Fig. S4 and S5\cite{SI}. In Fig. \ref{pd}, we plot the phase diagram of superconductivity in Te under pressure. Application of pressure effectively tunes both the crystal and electronic structure of Te. Te starts to become superconductive with $T_c$ of about 4.0 K by compressing it to about 4.1 GPa, accompanied by a structure transformation from Te-I  to Te-II. Then, with increasing pressure slightly, $T_c$ increases slowly in the narrow pressure region of Te-II. Te-II completely transforms to Te-III at around 8 GPa at room temperature. $T_c$ of Te-III is monotonically suppressed with external pressure, and $T_c$ can be suppressed to 2.6 K at around 30.7 GPa. With further increases in pressure above 34.8 GPa where Te-III transforms to Te-V, $T_c$ starts to increase rapidly and reaches a maximum value of 7.2 K at 34.8 GPa, followed by a decrease. It should be noted that Te exhibits a reversible superconducting state, which is in agreement with the high-pressure XRD results under decompression \cite{RN6,RN21}.\\

\indent The pressure dependence of the calculated band gaps of Te-I are shown in the upper panel of Fig. \ref{pd}. In Te-I, the nonmonotonic evolution of band gap is closely related to a pressure-induced topological transition\cite{agapito2013novel,hirayama2015weyl,deng2006unusual}. Then we calculate the DOS at Fermi surface for the high pressure phases, since $N(E_F)$ is a crucial parameter in BCS theory. It is observed that the DOS of Te-II forms a dome shape, while it decreases monotonically with increasing pressure in both Te-III and Te-V. This pressure dependence of the DOS agrees well with the variation of $T_c$ shown in the phase diagram.\\

\section{Conclusion}
To summarize, we have systematically investigated pressure-induced superconductivity in Te combining high-pressure $in$-$situ$ Raman spectroscopy, electrical transport, magnetic measurements and theoretical calculations. Under high pressure, Te shows multiple structural phase transitions with a nonmonotonic evolution of $T_c$. The superconducting phases of Te possess significantly different critical fields. The theoretical calculations demonstrate that the pressure dependence of the DOS agrees well with the variation of $T_c$. We present the superconducting phase diagram of Te and relate it with the sequential structural transitions. Our results will stimulate further studies on the interesting Te.\\

\begin{acknowledgments}
	\indent This work was supported by the National Natural Science Foundation of China (Grant Nos. 52272265, U1932217, 11974246, 12004252), the National Key R$\&$D Program of China (Grant No. 2018YFA0704300), and Shanghai Science and Technology Plan (Grant No. 21DZ2260400). The authors thank the support from Analytical Instrumentation Center ($\#$ SPST-AIC10112914), SPST, ShanghaiTech University.
\end{acknowledgments}     %IEEEtran为给定模板格式定义文件名

\nocite{*}
\bibliographystyle{apsrev4-2}
\bibliography{Te23}                        %ref为.bib文件名
\end{document}